\documentclass[12pt]{article}
\usepackage{amsfonts,amssymb,amsmath}
\usepackage[dvips]{epsfig}
\newcounter{tempeq}
\begin{document}
\title{\bf Generalized $su(2)$ coherent states for the Landau levels and their nonclassical properties }
\author{A. Dehghani$^{1,}$\thanks{Email: a\_dehghani@tabrizu.ac.ir,  alireza.dehghani@gmail.com}\hspace{2mm} and \hspace{2mm}
B. Mojaveri$^{2,}$\thanks{Email: bmojaveri@azaruniv.ac.ir; bmojaveri@gmail.com}  \\
{\small {\em $^1$Department of Physics, Payame Noor University, PO
Box 19395-4697, Tehran, Iran \,}}\\
{\small {\em $^2$Department of Physics, Azarbaijan Shahid Madani
University, PO Box 51745-406, Tabriz, Iran\,}}} \maketitle
\begin{abstract}
Following the lines of the recent papers [J. Phys. A: Math. Theor.
44, 495201 (2012); Eur. Phys. J. D 67, 179 (2013)], we construct
here a new class of generalized coherent states related to
the Landau levels, which can be used as the finite Fock subspaces
for the representation of the $su(2)$ Lie algebra. We establish the
relationship between them and the deformed truncated
coherent states. We have, also, shown that they satisfy the
resolution of the identity property through a positive definite
measures on the complex plane. Their nonclassical and quantum
statistical properties such as quadrature squeezing, higher
order `$su(2)$' squeezing, anti-bunching and anti-correlation
effects are studied in details. Particularly, the influence of the
generalization on the nonclassical properties of two modes is
clarified.
\\\\
 {\bf Keywords:} Nonlinear Coherent States, Sub-Poissonian Statistics, Squeezing Effect, Landau Levels.
\end{abstract}

\section{Introduction}
Coherent states (CSs), were first established by Schr\"{o}dinger
\cite{Schr¨odinger} as the eigenvectors of the boson annihilation
operator, $\hat{a}$, corresponding to the Heisenbereg-Weyl Lie
algebra. They play an important role in quantum optics and provide
us with a link between quantum and classical oscillators. Moreover,
these states can be produced by acting of the Glauber displacement
operator, $D(z)=e^{z \hat{a}^{\dag}-\overline{z}\hat{a}}$, on the
vacuum states, where $z$ is a complex variable. These states were
later applied successfully to some other models based on their Lie
algebra symmetries by Glauber \cite{Glauber1, Glauber2}, Klauder
\cite{Klauder1, Klauder2}, Sudarshan \cite{Sudarshan}, Barut and
Girardello \cite{Barut} and Perelomov \cite{Perelomov}.
Additionally, for the models with one degree of freedom either
discrete or continuous spectra- with no remark on the existence of a
Lie algebra symmetry- Gazeau et al proposed new CSs, which were
parametrized by two real parameters \cite{Gazeau1, Antoine}.
Moreover, there exist some considerations in connection with CSs
corresponding to the shape invariance symmetries \cite{Fukui,
Chenaghlou}. To construct CSs, four main different approaches the
so-called Schr\"{o}dinger, Klauder-Perelomov, Barut-Girardello, and
Gazeau-Klauder methods have been found, so that the second and the
third approaches rely directly on the Lie algebra symmetries and
their corresponding generators. Here, it is necessary to emphasize
that quantum coherence of states nowadays pervade many branches of
physics such as quantum electrodynamics, solid-state physics, and
nuclear and atomic physics, from both theoretical and experimental
viewpoints.

In addition to CSs, squeezed states (SSs) have attracted much
attention during past decades. These are non-classical states of the
electromagnetic field in which certain observables exhibit
fluctuations less than the vacuum state \cite{Stoler}. These states
are interesting because they can achieve lower quantum noise than
the zero-point fluctuations of the vacuum or coherent states. Over
the last four decades there have been several experimental
demonstrations of nonclassical effects, such as the photon
anti-bunching \cite{Kimble}, sub-Poissonian statistics \cite{Short,
Teich}, and squeezing \cite{Slusher, Wu}. Also, considerable
attention has been paid to the deformation of the harmonic
oscillator algebra of creation and annihilation operators
\cite{Biedenharn}. Some important physical concepts such as the CSs,
the even- and odd-CSs for ordinary harmonic oscillator have been
extended to deformation case. Moreover, there exist interesting
quantum interference effects related to the quantum states that are
namely superposition states, too \cite{Yurke, Noel}. Besides,
superpositions of CSs can be prepared in the motion of a trapped ion
\cite{Matos1, Monroe}. With respect to the nonclassical effects, the
coherent states turn out to define the limit between the classical
and nonclassical behavior.

 Another type of generalization of CSs is the nonlinear coherent
states (NLCSs), or f-CSs \cite{Solomon}. They are associated with
nonlinear algebras and defined as the eigenstates of the
annihilation operator of a f-deformed oscillator $f(\hat{N})a$.
Indeed, the nature of the nonlinearity depends on the choice of the
function $f(\hat{N})$ \cite{Manko}. These states may appear as
stationary states of the center-of mass motion of a trapped ion
\cite{Matos2, Raffa}. NLCSs exhibit nonclassical features such as
quadrature squeezing, sub-Poissonian statistics, anti-bunching,
self-splitting effects and so on [28-30].

The discrete energy values corresponding to the motion of a charged
particle on a flat surface in the presence of a uniform external
magnetic field perpendicular to this plane are called Landau levels
\cite{Landau}. The physics of charged particles in the presence of
magnetic field has been one of the important problems in quantum
mechanics, inspired by condensed matter physics, quantum optics etc.
From the classical viewpoint, the magnetic field creates a
transverse current perpendicular to both the direction of motion of
particle and the direction of the magnetic field. Therefore, the
Landau problem can be counted as a cornerstone of quantum Hall
effect. In the last three decades, many efforts have been carried
out to describe the spectral properties of the quantum Hall effect
[32-39]. The quantum Hall effect, as a universal phenomenon, is
observed on any two-dimensional surface with a charged particle
moving in the presence of a strong perpendicular uniform magnetic
field. In metals and other dense electronic systems the electrons
occupy many Landau levels. Furthermore, the kinetic energy levels of
electrons in two-dimensional gas correspond to Landau levels. For
all these reasons, the study of coherent states for Landau levels is
of great importance which was first studied by Malkin and Man'ko
\cite{Malkin1} and later by many others [41-55].

 Recently, we have introduced the generalized coherent states (GCSs) for harmonic and pseudo harmonic
oscillators based on the generalization of the bosonic displacment
operators associated with the Heisenbereg-Weyl and $su(1,1)$ Lie
algebras, respectively \cite{Dehghani,Mojaveri6}. It has been shown
that, they are new class of nonlinear coherent states and have some
interesting features such as temporal stability and nonclassical
properties. An interesting feature of the above mentioned approach
is due to the fact that, contrary to the Klauder-Perelomov and
Barut-Girardello approaches, they do not require for existence of
dynamical symmetries associated with the considered system. In the
other word, we need only the raising operator associated with the
considered system in the framework of supersymmetric quantum
mechanics.

 Along with extension of the above scenarios to the $su(2)$ Lie algebra counterpart,
we introduce generalized $su(2)$ CSs for the problem of an
electron moving in the constant magnetic field . These states admit
a resolution of the identity through positive definite and
non-oscillating measures on the complex plane. We have shown that
these states are nonlinear truncated coherent states with a spacial
nonlinearity function. It has been discussed, in details, that they
have indeed nonclassical features such as squeezing effect and
sub-Poissonian statistics. The results are compared with the
properties of the well known $su(2)$ coherent states \cite{Buzek3}.

 This paper is organized as follows: in section 2, we briefly review
on the Schwinger representation of the $su(2)$ Lie algebra symmetry
of Landau levels. Section 3 is devoted to construction the new class
of generalized $su(2)$ CSs $|z\rangle_{r}$, via generalized
analogue of the displacement operators acting on the Landau levels
with lowest z-angular momentum. In order to realize the resolution
of the identity, we have found the positive definite measures on the
complex plane. There, it has been shown that these states can be
interpreted as the nonlinear truncated coherent states. Furthermore,
in section 4 by evaluating some physical quantities, we discuss
their statistical and nonclassical properties. Finally, we conclude
the paper in section 5.

\section{Landau levels}
Let us first explain the exact solvability of the symmetric-gauge
Landau Hamiltonian corresponding to the motion of an electron on a
flat surface in the presence of the uniform magnetic field in the
positive direction of $z$-axis, i.e.
$H=\hbar\omega(a^{\dagger}a+1/2)=\hbar\omega(b^{\dagger}b+1/2)-\omega
L_3$ with $L_3=-i\hbar\partial/\partial\varphi$. It has an
infinite-fold degeneracy on the Landau levels, that is
\setcounter{equation}{0}
\begin{eqnarray}
H\left|n,m\right\rangle=\hbar\omega(n+1/2)\left|n,m\right\rangle,
\end{eqnarray}
in which Landau cyclotron frequency is expressed in terms of the
value of the electron charge, its mass, the magnetic field strength
$B$ and also the velocity of light as $\omega=eB/Mc$. Here, $m$ is
an integer number and $n$ is a nonnegative one together with
$n\geq-m$ limitation. Each pair of operators $(a,a^{\dagger})$ and
$(b,b^{\dagger})$ have the following explicit forms in terms of the
polar coordinates $0<r<\infty$ and $0\leq\varphi<2\pi$ for
two-dimensional flat surface \cite{Mojaveri5},
\begin{eqnarray}
&&\hspace{-17mm}a=-e^{i\varphi}\sqrt{\frac{\hbar}{2M\omega}}\left(\frac{\partial}{\partial
r}+\frac{i}{r}\frac{\partial}{\partial\varphi}+\frac{M\omega}{2\hbar}r\right),\hspace{5.5mm}
a^{\dagger}=e^{-i\varphi}\sqrt{\frac{\hbar}{2M\omega}}\left(\frac{\partial}{\partial
r}-\frac{i}{r}\frac{\partial}{\partial\varphi}-\frac{M\omega}{2\hbar}r\right),\\
&&\hspace{-17mm}b=e^{-i\varphi}\sqrt{\frac{\hbar}{2M\omega}}\left(\frac{\partial}{\partial
r}-\frac{i}{r}\frac{\partial}{\partial\varphi}+\frac{M\omega}{2\hbar}r\right),\hspace{7mm}
b^{\dagger}=-e^{i\varphi}\sqrt{\frac{\hbar}{2M\omega}}\left(\frac{\partial}{\partial
r}+\frac{i}{r}\frac{\partial}{\partial\varphi}-\frac{M\omega}{2\hbar}r\right),
\end{eqnarray}
and form two separate copies of Weyl-Heisenberg algebra,
\begin{eqnarray}\label{WeHeAl}
[a,a^{\dagger}]=1,\,\,\,\,[b,b^{\dagger}]=1,\,\,\,\,[a,b^{\dagger}]=[a^{\dagger},b]=[a,b]=[a^{\dagger},b^{\dagger}]=0,
\end{eqnarray}
with the unitary representations as
\begin{eqnarray}\label{arep}
&&a\left|n,m\right\rangle=\sqrt{n}\left|n-1,m+1\right\rangle,
\hspace{16mm}
a^{\dagger}\left|n-1,m+1\right\rangle=\sqrt{n}\left|n,m\right\rangle,\\\label{brep}
&&b\left|n,m\right\rangle=\sqrt{n+m}\left|n,m-1\right\rangle,
\hspace{15mm}
b^{\dagger}\left|n,m-1\right\rangle=\sqrt{n+m}\left|n,m\right\rangle.
\end{eqnarray}
Infinite fold degeneracy via the magnetic quantum number $m$ is a
result of existence of rotational symmetry, i.e. $[H,L_3]=0$.
Besides, the infinite Fock subspace corresponding to a given energy
represents the dynamical symmetry group of Weyl-Heisenberg:
$[H,b]=[H,b^{\dagger}]=0$. Landau levels are orthonormal with
respect to integration over the entire plane, that is
\begin{eqnarray}\label{ortho}
&&\left\langle
n,m|n^{\prime},m^{\prime}\right\rangle:=\int_{\varphi=0}^{2\pi}\int_{r=0}^{\infty}\left\langle
r,\varphi|n,m\right\rangle^{*} \left\langle
r,\varphi|n^{\prime},m^{\prime}\right\rangle r dr d\varphi
=\delta_{n n^{\prime}}\delta_{m m^{\prime}},
\end{eqnarray}
in which
\begin{eqnarray}
&&\left\langle
r,\varphi|n,m\right\rangle=\sqrt{\frac{n!}{\pi(n+m)!}\left(\frac{M\omega}{2\hbar}\right)^{m+1}}r^m\,
e^{im\varphi} e^{\frac{-M\omega}{4\hbar}r^2} L_n^{(m)}(M\omega
r^2/2\hbar)
\end{eqnarray}
is the polar coordinate representation of Landau levels in terms of
the associated Laguerre functions. At the end of this section, let
us denote the Hilbert space corresponding to all Landau levels with
$\mathcal{H}=span\{\left|n,m\right\rangle\}_{n=0,\hspace{1mm}m=-n}^{\infty\hspace{5mm}\infty}$.
For a given $n$ the Hilbert space $\mathcal{H}$ can be split into
the infinite direct sums of finite dimensional Hilbert subspaces
$\mathcal{H}_{n}$, i.e.
$\mathcal{H}=\oplus_{n=0}^{\infty}\mathcal{H}_{n}$ with
$\mathcal{H}_n=span\{\left|n-k,-n+2k\right\rangle\}_{k=0}^{n}$
\cite{Mojaveri5}.

\section{Generalized $su(2)$ CSs for Landau levels}
Using the commutation relations of the Weyl-Heisenberg algebras
given in Eq.(4), one can easily find that the following generators
\setcounter{tempeq}{\value{equation}}
\renewcommand\theequation{\arabic{tempeq}\alph{equation}}
\setcounter{equation}{-1} \addtocounter{tempeq}{1}
\begin{eqnarray}
&&\hspace{-1.5cm}
{K}_{-}:=a^{\dagger}b,\hspace{1cm}{K}_{+}:=ab^{\dagger},\hspace{1cm}K_{3}:=\frac{b^{\dagger}b-a^{\dagger}a}{2},
\end{eqnarray}
produce the bosonic and unitary representation of the $su(2)$ Lie
algebra with the standard commutation relations
\setcounter{tempeq}{\value{equation}}
\renewcommand\theequation{\arabic{tempeq}\alph{equation}}
\setcounter{equation}{-1} \addtocounter{tempeq}{10}
\begin{eqnarray}
&&\hspace{-1.5cm}\left[{K}_{+},{K}_{-}\right]=2{K}_{3},
\hspace{20mm}\left[{K}_{3},{K}_{\pm}\right]=\pm{K}_{\pm}.
\end{eqnarray}
From Eqs. (5) and (6), it becomes obvious that the finite
dimensional Hilbert space $\mathcal{H}_n$ represent the $su(2)$ Lie algebra as
\setcounter{tempeq}{\value{equation}}
\renewcommand\theequation{\arabic{tempeq}\alph{equation}}
\setcounter{equation}{0} \addtocounter{tempeq}{11}
\begin{eqnarray}
&&\hspace{-15mm}{K}_{+}|n+1,m-2\rangle=\sqrt{(n+1)\left(n+m\right)}\,|n,m\rangle,\\
&&\hspace{-15mm}{K}_{-}|n,m\rangle=\sqrt{(n+1)\left(n+m\right)}\,|n+1,m-2\rangle,\\
&&\hspace{-15mm}{K}_{3}|n,m\rangle=\frac{m}{2}\,|n,m\rangle.
\end{eqnarray}
Note that ${K}_{3}$ is a self-adjoint operator and the two
operators ${K}_{-}$ and ${K}_{+}$ are Hermitian conjugate of each
other with respect to the inner product (7).

 Now, generalized $su(2)$ CSs for an electron in the constant magnetic field are introduced via generalized analogue of the
displacement operators acting on the Landau levels with lowest
z-angular momentum, $|n,-n\rangle$, as
\setcounter{tempeq}{\value{equation}}
\renewcommand\theequation{\arabic{tempeq}\alph{equation}}
\setcounter{equation}{-1} \addtocounter{tempeq}{9}
\begin{eqnarray}
&&\hspace{-15mm}|z\rangle_{r}:=
{M_{r}}^{-\frac{1}{2}}(|z|)_{1}F_{r}\left([n-1] ,[n-1,n,n+1,
...,n+r-1],z{K}_{+} \right)|n,-n\rangle, \hspace{5mm} r\geq1,
\end{eqnarray}
where $z$$(= |z|e^{i\phi})$ and $r$ are respectively the coherence
and the deformation parameters, respectively. Clearly,
$|z\rangle_{r}$ becomes the standard $su(2)$ CSs corresponding to
bosonic realization of the $su(2)$ Lie algebra \cite{Buzek3}, when
$r$ tends to unity and $z$ be replaced with
$\frac{z}{|z|}\tanh(|z|)$. Using by the series form of the
hypergeometric functions and applying the Eq. (11(a)), the states
$|z\rangle_{r}$ can be obtained as
\setcounter{tempeq}{\value{equation}}
\renewcommand\theequation{\arabic{tempeq}\alph{equation}}
\setcounter{equation}{-1} \addtocounter{tempeq}{13}
\begin{eqnarray}
&&\hspace{-15mm}|z\rangle_{r}={M_{r}}^{-\frac{1}{2}}(|z|)
\sum_{k=0}^{n}z^{k}\left[\prod_{l=1}^{r-1}{\frac{\Gamma(n+l-1)}{\Gamma(n+k+l-1)}}\right]
\sqrt{\frac{n!}{(n-k)!k!}}|n-k,-n+2k\rangle, \hspace{2mm} r\geq2,
\end{eqnarray}
where the normalization constant ${M_{r}}(|z|)$ is chosen so that
$|z\rangle_{r}$ is normalized to unity with respect to the inner
product (7), i.e. $_{r}\langle z|z\rangle_{r}=1$, then
\setcounter{tempeq}{\value{equation}}
\renewcommand\theequation{\arabic{tempeq}\alph{equation}}
\setcounter{equation}{-1} \addtocounter{tempeq}{14}
\begin{eqnarray}
&&\hspace{-15mm}{M_{r}}(|z|)=_{1}F_{2r-2}\left(\left[-n\right],
\left[n , n, ..., n+r-2, n+r-2\right],-|z|^2\right).
\end{eqnarray}
It should be noticed that, these states can be categorized as
special class of {\em {Generalized Hypergeometric CSs}} \cite{Appl}
have already been made by Appl et al.  Using the inner product (7),
the overlapping of the generalized $su(2)$ CSs can be
calculated as follows: \setcounter{tempeq}{\value{equation}}
\renewcommand\theequation{\arabic{tempeq}\alph{equation}}
\setcounter{equation}{0} \addtocounter{tempeq}{15}
\begin{eqnarray}
&&\hspace{-15mm}_{r}\langle z_{1}|z_{2}\rangle_{r}=
\frac{_{1}F_{2r-2}\left(\left[-n\right],\left[n, n, ..., n+r-2, n+r-2\right],-\overline{z}_{1}z_{2}\right)}{\sqrt{{M_{r}}(|z_{1}|){M_{r}}(|z_{2}|)}},\\
&&\hspace{-15mm}_{r_{1}}\langle z|z\rangle_{r_{2}}=
\frac{_{1}F_{r_{1}+r_{2}-2}\left(\left[-n\right],\left[n, n+1, ...,
n+r_{1}-2,
n+r_{2}-2\right],-|z|^2\right)}{\sqrt{{M_{r_{1}}}(|z|){M_{r_{2}}}(|z|)}},
\end{eqnarray}
and result that two different kinds of these states are
non-orthogonal, if $r_{1}\neq r_{2}, z_{1}\neq z_{2}$. Now, we
should check realization of the property of \textit{resolution to
identity} for the states $|z\rangle_{r}$ in the Hilbert space
$\mathcal{H}_n$.
 \setcounter{tempeq}{\value{equation}}
\renewcommand\theequation{\arabic{tempeq}\alph{equation}}
\setcounter{equation}{-1} \addtocounter{tempeq}{14}
\begin{eqnarray}
&&\hspace{-13mm}1_{\mathcal{H}} = \oint_{\mathbb{C}(z)}d^{2}z
K_{r}(|z|){|z\rangle_{r}\,{_{r}\langle z|}}=
\sum_{m=-n}^{n}|n,m\rangle\langle n,m|,
\end{eqnarray}
where $K_{r}(|z|)$ is a positive definite measure to be determined below.
By substituting Eq. (13) in Eq. (16) lead us to the following
integral relation \setcounter{tempeq}{\value{equation}}
\renewcommand\theequation{\arabic{tempeq}\alph{equation}}
\setcounter{equation}{-1} \addtocounter{tempeq}{17}
\begin{eqnarray}
&&\hspace{-13mm} \int_{0}^{\infty}d|z||z|^{2k+1}\frac{\pi
K_{r}(|z|)}{M_{r}(|z|)}}={\left[\prod_{l=1}^{r-1}{\frac{\Gamma(n+k+l-1)}{\Gamma(n+l-1)}}\right]^{2}
{\frac{(n-k)!k!}{n!}}.
\end{eqnarray}
Along with the integral relation for the Meijers G-functions (see
$\frac{7-811}{4}$ in \cite{Gradshteyn}), the positive definite and
non-oscillating measure $K_{r}(|z|)$ is obtained as
\setcounter{tempeq}{\value{equation}}
\renewcommand\theequation{\arabic{tempeq}\alph{equation}}
\setcounter{equation}{-1} \addtocounter{tempeq}{18}
\begin{eqnarray}
&&\hspace{-15mm}K_{r}(|z|)=\frac{2\,\,_{1}F_{2r-2}\left(\left[-n\right],\left[n,
, n, ..., n+r-2, n+r-2\right],|z|^2\right)}{\pi
\left[\prod_{l=1}^{r-1}\Gamma(n+l-1)\right]^{2}}\nonumber\\
&&\hspace{4mm}\times G\left(\left[[-n-1],[\hspace{1mm}]], [[0, n-1,
n-1, ..., n+r-3, n+r-3],[\hspace{1mm}]\right],|z|^2\right).
\end{eqnarray}
\begin{figure}
\centering
\includegraphics[width=500 pt]{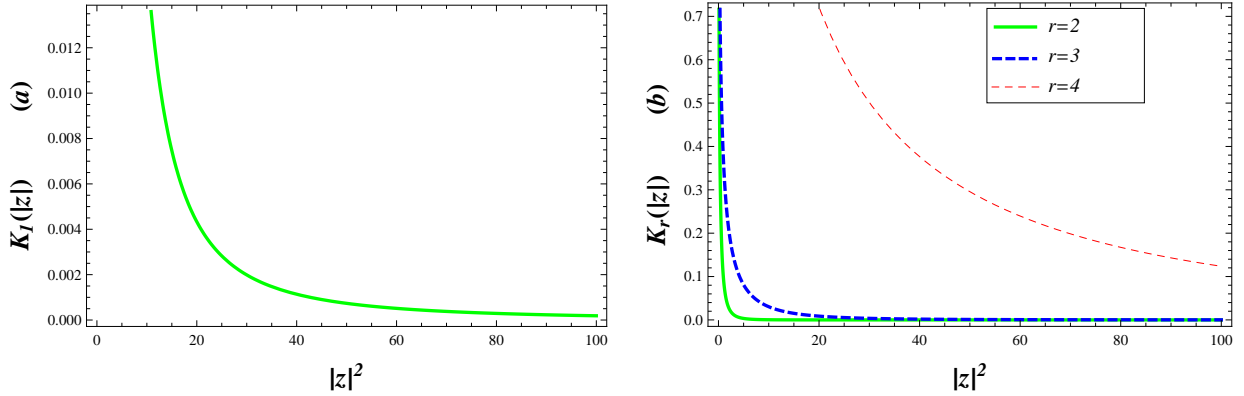}
\caption{Plots of the positive definite measures $K_{r}(|z|)$ in
terms of $|z|^2$ for different values of $r$. $(a)$ relates to $r=1$
and $(b)$ is corresponds to higher amounts of $r=2,3$ and 4.}
\end{figure}
For ($r=1,2,3$ and 4) we have plotted the changes of the
non-oscillating and positive definite measures $K_{r}(|z|)$ in terms
of $|z|^2$ in Figures 1.\\\\
\subsection{Generalized $su(2)$ CSs as the deformed truncated CSs}
Now, we show that the states $|z\rangle_{r}$ can be
interpreted as deformed truncated CSs associated with $f$-deformed
Schwinger realization of $su(2)$ Lie algebra (10) with a special
deformation function. The deformed truncated CSs \cite{Mahdifar1,
Mahdifar2} are defined, in the similar approach of constructing of
the truncated CSs \cite{Kuang}, as
\setcounter{tempeq}{\value{equation}}
\renewcommand\theequation{\arabic{tempeq}\alph{equation}}
\setcounter{equation}{-1} \addtocounter{tempeq}{19}
\begin{eqnarray}
\left|z,f\right\rangle=C^{\,-\frac{1}{2}}(|z|)\,\,exp\left(z\,f(\hat{n_a},\hat{n_b})ab^{\dagger}\right)\left|n,-n\right\rangle,
\end{eqnarray}
where $f(\hat{n_a},\hat{n_b})$ is the deformation function which
depends on the photon numbers of $\hat{n_1}$ and $\hat{n_2}$. Now by
using the following relations
\setcounter{tempeq}{\value{equation}}
\renewcommand\theequation{\arabic{tempeq}\alph{equation}}
\setcounter{equation}{0} \addtocounter{tempeq}{20}
\begin{eqnarray}
&&\hspace{-15mm}a f(\hat{n}_a,\hat{n}_b)=f(\hat{n}_a+1,\hat{n}_b)a,\\
&&\hspace{-15mm}b^{\dagger}f(\hat{n}_a,\hat{n}_b)=f(\hat{n}_a,\hat{n}_b-1)b^{\dagger},
\end{eqnarray}
we obtain the expansion of the $\left|z,f\right\rangle$ as
\renewcommand\theequation{\arabic{equation}}
\setcounter{equation}{\value {tempeq}}
\begin{eqnarray}
&&\hspace{-15mm}\left|z,f\right\rangle=C^{\,-\frac{1}{2}}(|z|)\sum_{k=0}^{n}z^k
\left[f(n_a+k-1,n_b-k+1)\right]!\,\sqrt{\frac{n!}{k!(n-k)!}}\left|n-k,-n+2k\right\rangle
\end{eqnarray}
where $\left[f(n_a+k-1,n_b-k+1)\right]!=
\prod_{s=0}^{k-1}\,f(n_a+s,n_b-s)$. Now, by comparing the Eq. (13)
with the Eq. (21), one can find that $\left|z\right\rangle_{r}$ can
be considered as deformed truncated CSs with the deformatin
function
$f(\hat{n}_a,\hat{n}_b)=\frac{\Gamma(-\hat{n}_a+2n-1)}{\Gamma(-\hat{n}_a+2n+r-2)}$.
Obviously, it tends to unity for $r = 1$.

\section{Statistical properties of generalized $su(2)$ CSs}
In this section, some quantum statistical properties
including squeezing in field quadratures ($x$ and
it's canonical conjugate $p_{x}$), higher order squeezing, photon
statistics and cross-correlation of these states are studied in
details. They provide appropriate framework for the study of
(non-)classical features of these quantum
states \cite{Shchukin}.\\\\
$\diamondsuit${\em{Squeezing in Field Quadratures
$x$ and $p_{x}$}}\\
To investigate the quadrature squeezing in the introduced
generalized CSs we now construct the creation and annihilation
operators of the system in terms of the coordinates $x$ and its
conjugate momenta $p_x$. From Eqs. (2) and (3), the operators $x$
and $p_x$ can be easily written as
\begin{eqnarray}
&&\hspace{-34mm}x=\sqrt{\frac{\hbar}{2M\omega}}(b+b^{\dagger}-a-a^{\dagger}),\hspace{1.3cm}
p_x=\frac{i}{2}\sqrt{\frac{M\hbar\omega}{2}}(b-b^{\dagger}+a-a^{\dagger}),
\end{eqnarray}
which satisfy the commutation relation $[x,p_x]=i\hbar$.  From
these, the uncertainty condition for the variances of the
quadratures $x$, $p_x$ follow
\begin{eqnarray}
&&\hspace{-34mm}\sigma_{xx}\sigma_{p_xp_x}\geq \frac{\hbar^2}{4},
\end{eqnarray}
where $\sigma_{ab}=\frac{1}{2}<ab+ba>-<a><b>$. To characterize the
degree of squeezing in the $x$ and $p_x$ components, we introduce
the following parameters \cite{Mojaveri7}
\begin{eqnarray}
&&\hspace{-34mm}S_x=\frac{2\sigma_{xx}}{\sqrt{\hbar^2+4\sigma_{xp_x}^2}}-1,\hspace{2cm}
S_{p_x}=\frac{2\sigma_{p_xp_x}}{\sqrt{\hbar^2+4\sigma_{xp_x}^2}}-1.
\end{eqnarray}
\begin{figure}
\centering
\includegraphics[width=400 pt]{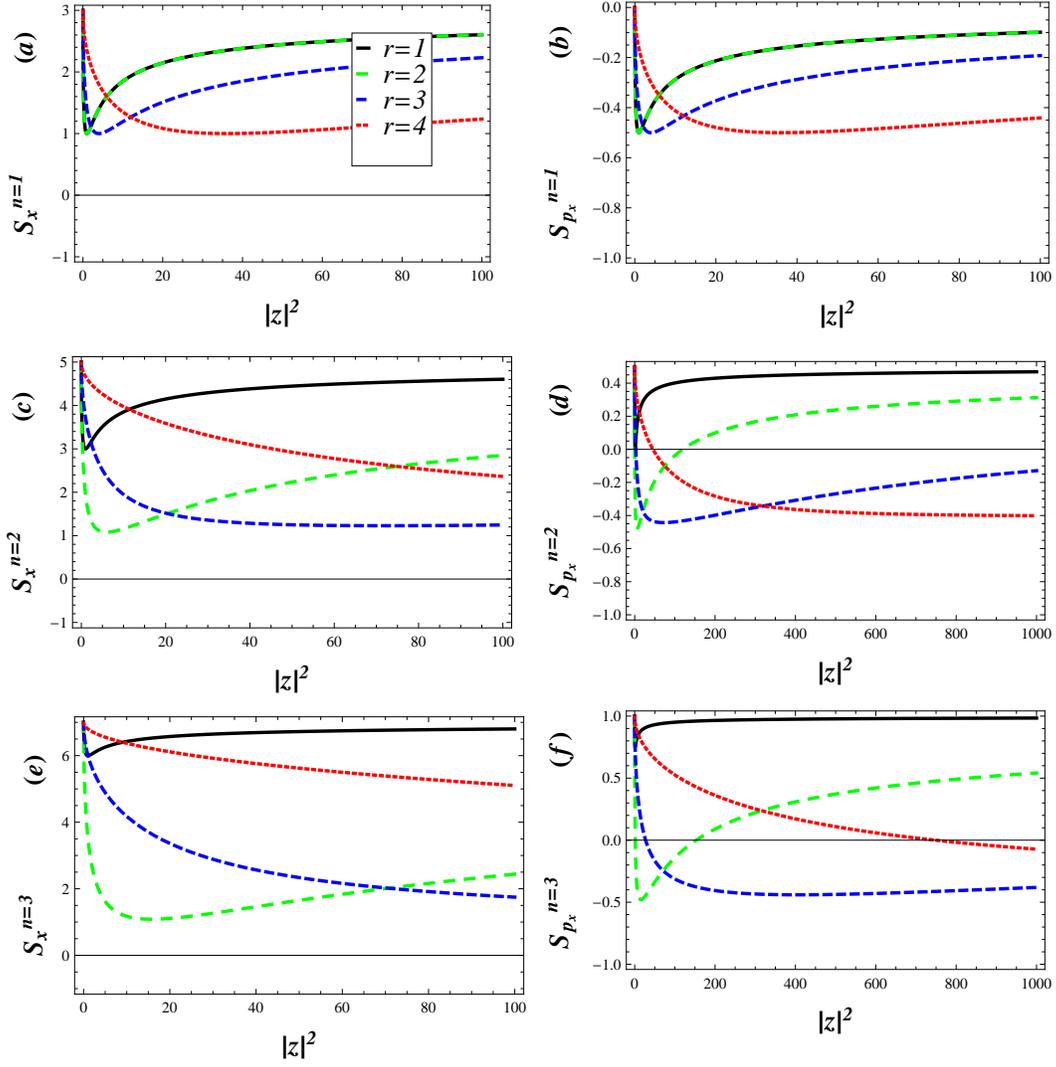}
\caption{The plots of quadrature squeezing parameter $ S_{x}^{n}$
($a,c,e$) and $ S_{p_x}^{n}$ $(b,d,f)$ respectively, against $|z|^2$
for $\phi=0$ and different values of $n$ and $r$.}
\end{figure}
\begin{figure}
\centering
\includegraphics[width=400 pt]{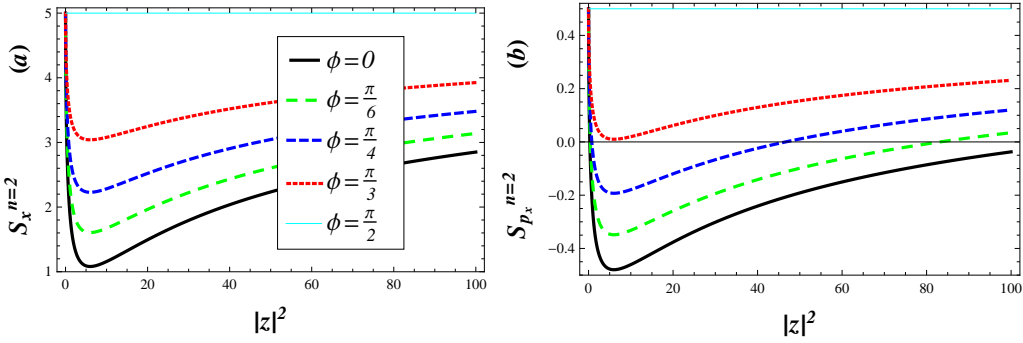}
\caption{The plots of quadrature squeezing parameter $ S_{x}^{n}$
($a$) and $ S_{p_x}^{n}$ $(b)$ respectively, against $|z|^2$ for
different values of $\phi$ by setting $r=2$ and $n=2$.}
\end{figure}
For our generalized $su(2)$ CSs (Eq. (13)) we have
\begin{eqnarray}
&&\hspace{-34mm}\langle a \rangle=\langle b \rangle=\langle a^2
\rangle=\langle b^2 \rangle=\langle a b \rangle=0,
\end{eqnarray}
from which it follows that \setcounter{tempeq}{\value{equation}}
\renewcommand\theequation{\arabic{tempeq}\alph{equation}}
\setcounter{equation}{0} \addtocounter{tempeq}{1}
\begin{eqnarray}
&&\hspace{-15mm}\left(\frac{M\omega}{\hbar}\right)S_{x}=1+n-2|z|\nonumber\\
&&\hspace{-25mm}\times\frac{_{1}F_{2r-2}\left([1-n]
,[n,n+1,n+1,...,n+r-2,n+r-2,n+r-1],-|z|^2\right)}{_{1}F_{2r-2}\left([-n]
,[n,n,n+1,n+1, ...,n+r-2,n+r-2],-|z|^2\right)}\cos\phi,\\
&&\hspace{-15mm}\left(\frac{M\omega}{\hbar}\right)S_{p_{x}}=1+n-2|z|\nonumber\\
&&\hspace{-25mm}\times\frac{_{1}F_{2r-2}\left([1-n]
,[n,n+1,n+1,...,n+r-2,n+r-2,n+r-1],-|z|^2\right)}{_{1}F_{2r-2}\left([-n]
,[n,n,n+1,n+1, ...,n+r-2,n+r-2],-|z|^2\right)}\cos\phi.
\end{eqnarray}
Quadrature squeezing in the generalized $su(2)$ CSs exists if $S_x$
or $S_{p_x}$ is in the range $(-1,0)$. It follows from Eqs. (26)
that the quantities $S_x$ and $S_{p_x}$ are dependent on the Landau
cyclotron frequency $\omega$ which comes from the variations of
magnetic field $B_{ext}$ too. In Figure 2, we show $S_x$ and
$S_{p_x}$ as functions of $|z|^{2}$ for $r$(= 1, 2, 3 and 4) as well
as $n$(= 1,2 and 3) for fixed $(\phi=0,M=\hbar=\omega=1)$. From
this, there is no squeezing in the $x$ component. As shown, for
$r=1$ quadrature squeezing in the $p_x$ component occurs only for
$n=1$. In the other word the well-known $su(2)$ CSs for Landau
levels do not exhibit quadrature squeezing for $n>1$ while, our
introduced states have squeezing property for any values of $n$.
Meanwhile, from Figure 2 $(b,d,f)$, by increasing the generalization
parameter $r$, quadrature squeezing hold for large value of $|z|^2$.
Also, we show in Figure 3 the squeezing parameter $S_{x}$ and
$S_{p_x}$ for different values of $\phi$(= 0, $\frac{\pi}{6}$,
$\frac{\pi}{4}$, $\frac{\pi}{3}$, and $\frac{\pi}{2}$) for fixed
$(r=2,M=\hbar=\omega=1)$. This figure
show that quadrature squeezing in the $p_x$ component occur for $\phi< \frac{\pi}{3}$.\\\\
$\diamondsuit${\em{Higher Order `$su(2)$' Squeezing}}\\
In order to study $su(2)$ squeezing we consider the
following Hermitian quadrature operators
\renewcommand\theequation{\arabic{equation}}
\setcounter{equation}{\value {tempeq}}
\begin{eqnarray}
&&\hspace{-14mm}X_{1}=\frac{{K}_{+}+{K}_{-}}{2},
\hspace{4mm}X_{2}=\frac{{K}_{-}-{K}_{+}}{2i},
\end{eqnarray}
with the commutation relation $[X_{1}, X_{2}] = -i{K}_{3}$. From
this communication relation the uncertainty relation for the
variances of the quadrature operators $X_{i}$ follows
\begin{eqnarray}
&&\hspace{-14mm}\langle(\Delta X_{1})^2\rangle \langle(\Delta
X_{2})^2\rangle \geq \frac{|\langle{K}_{3}\rangle|^{2}}{4},
\end{eqnarray}
where $\langle(\Delta X_{1})^2\rangle= \langle
\left(X_{1}\right)^2\rangle-{\langle X_{1}\rangle}^2$ and the
angular brackets denote averaging over an arbitrary normalizable
state for which the mean values are well defined, $\langle
X_{i}\rangle={_{r}\langle z|}X_{i}|z\rangle_{r}$. Following Walls
(1983) as well as Wodkiewicz and Eberly (1985) \cite{Walls,
WODKIEWICZ} we will say the state is $su(2)$ squeezed if the
variance of one of the generators $X_i$ be smaller than the average
uncertainty of the quantum-mechanical fluctuations:
\begin{eqnarray}
&&\hspace{-14mm}\langle(\Delta X_{i})^2\rangle <
\frac{|\langle{K}_{3} \rangle|}{2},\hspace{4mm} for\hspace{2mm} i=1
\hspace{2mm}or \hspace{2mm}2.
\end{eqnarray}
To measure the degree of the $su(2)$ squeezing we introduce the
squeezing factor $S_{i}$ \cite{Buzek}
\begin{eqnarray}
&&\hspace{-14mm}S_{i}=\frac{\langle(\Delta
X_{i})^2\rangle-\frac{|\langle{K}_{3}
\rangle|}{2}}{\frac{|\langle{K}_{3} \rangle|}{2}},
\end{eqnarray}
it leads that the $su(2)$ squeezing condition takes on the simple
form $S_{i}< 0$, however maximally squeezing is obtained for
$S_{i}=-1$. By using of the mean values of the generators of the
$su(2)$ Lie algebra, one can derive the variance of the quadrature
operators $X_{1(2)}$ as \setcounter{tempeq}{\value{equation}}
\begin{eqnarray}
&&\hspace{-14mm}\langle(\Delta X_{1(2)})^2\rangle
=\frac{2\left\langle{K}_{+}{K}_{-}\right\rangle-2\left\langle{K}_{3}\right\rangle\pm\left\langle{{K}_{+}}^{2}+{{K}_{-}}^{2}\right\rangle-
{\left\langle{{K}_{-}}\pm{{K}_{+}}\right\rangle}^{2}}{4}.
\end{eqnarray}
With the help of following mean values of the generators of the
$su(2)$ Lie algebra with respect to states (13)
\begin{figure}
\centering
\includegraphics[width=450 pt]{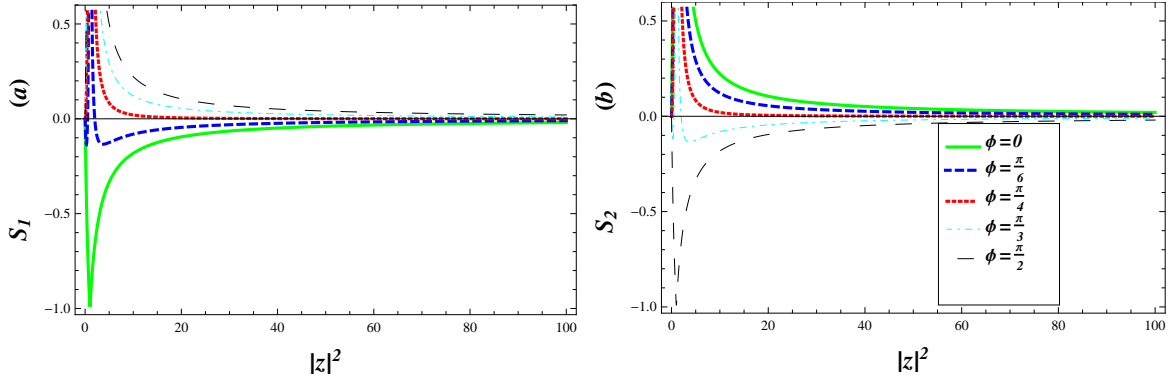}
\caption{The plots of $su(2)$ squeezing parameter (a) $S_1$ and (b)
$S_2$ versus $|z|^2$ for different values of $\phi$ and $n=1$ for
the standard $su(2)$ CSs for Landau levels.}
\end{figure}
\begin{figure}
\centering
\includegraphics[width=350 pt]{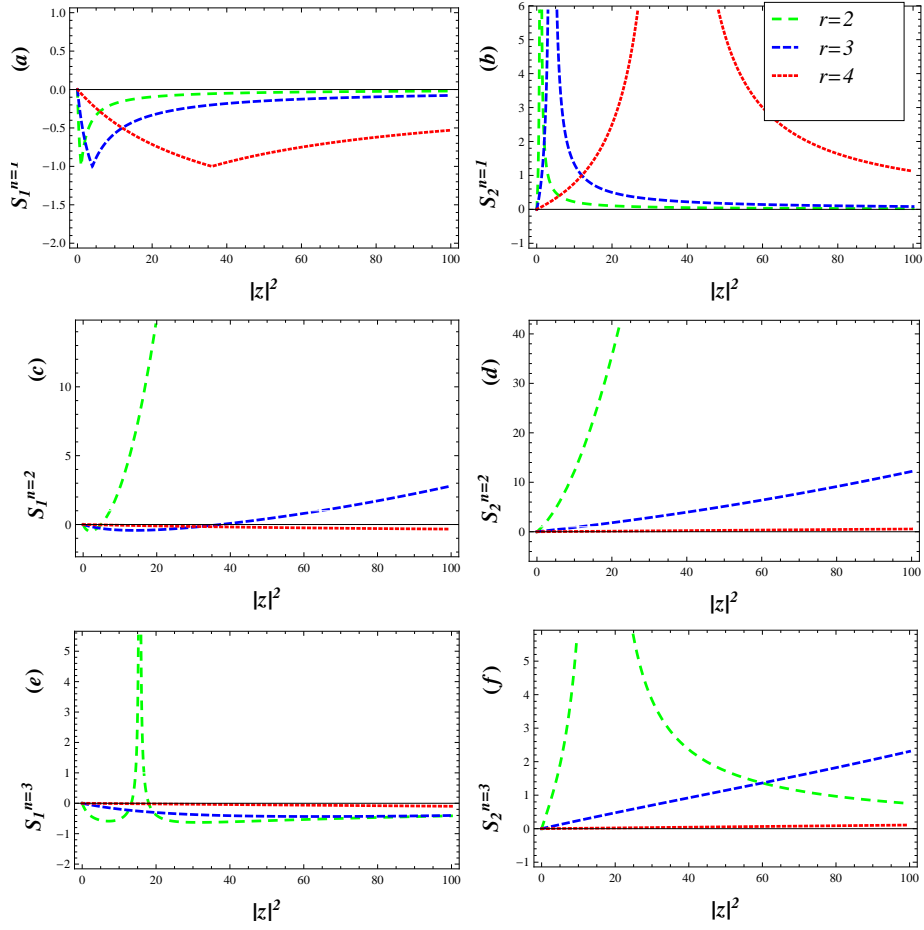}
\caption{The plots of quadrature squeezing parameter $ S_{1}^{n}$
($a,c,e$) and $ S_{2}^{n}$ $(b,d,f)$, respectively versus $|z|^2$
for different values of $r$ as well as different $n$ while we choose
the phase $\phi=0$.}
\end{figure}
\begin{figure}
\centering
\includegraphics[width=350 pt]{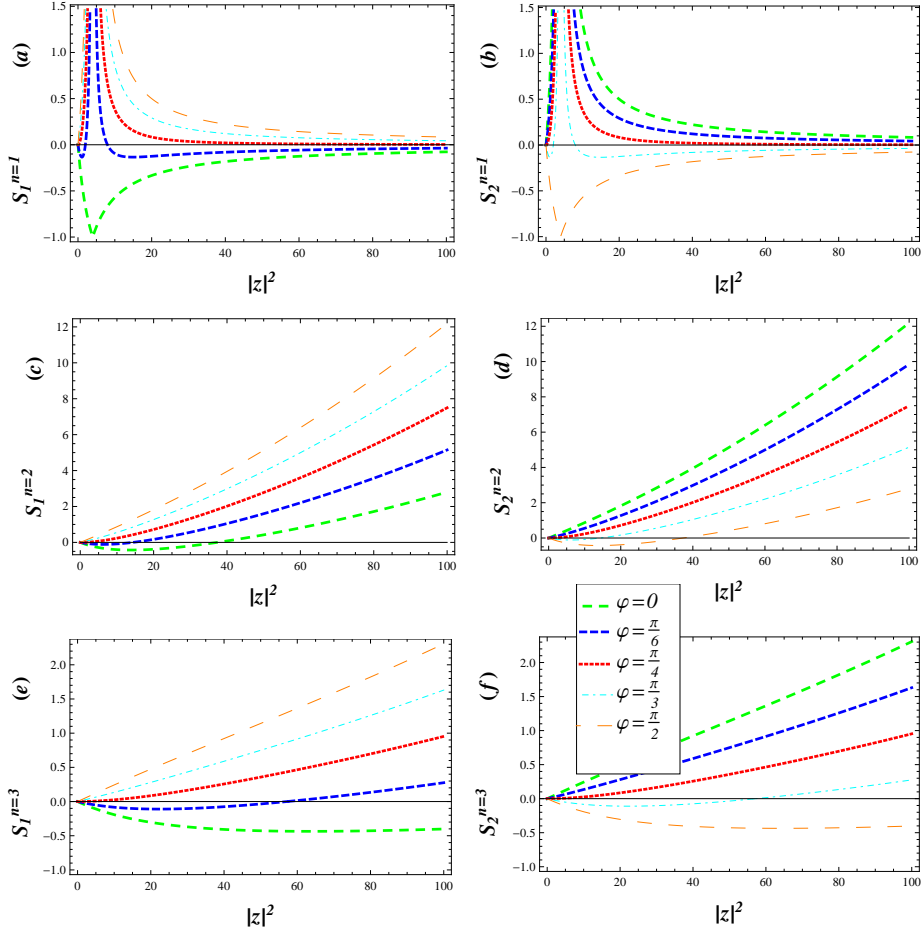}
\caption{The plots of quadrature squeezing parameter $ S_{1}^{n}$
($a,c,e$) and $ S_{2}^{n}$ $(b,d,f)$ respectively, against $|z|^2$
for $r=4$ and different values of $n$ and $\phi$.}
\end{figure}
\setcounter{tempeq}{\value{equation}}
\renewcommand\theequation{\arabic{tempeq}\alph{equation}}
\setcounter{equation}{-1} \addtocounter{tempeq}{27}
\begin{eqnarray}
&&\hspace{-20mm}\left\langle{K}_{+}\right\rangle=\overline{\left\langle{K}_{-}\right\rangle}=
\frac{\Gamma(n+1)}{\Gamma(n+r-1)}\overline{z}\nonumber\\
&&\hspace{-18mm}\times\frac{{_{1}F_{2r-2}\left(\left[1-n\right],\left[n,n+1,n+1,
..., n+r-2,n+r-2,n+r-1\right],-|z|^{2}\right)}}{_{1}F_{2r-2}\left(\left[-n\right],\left[n,n, ..., n+r-2,n+r-2\right],-|z|^2\right)},\nonumber\\
&&\hspace{-20mm}\left\langle{{K}_{+}^{\lambda}}^{2}\right\rangle=\overline{\left\langle{{K}_{-}^{\lambda}}^{2}\right\rangle}=\left(\frac{n-1}{n+r-1}\right)\left(\frac{\Gamma(n+1)}
{\Gamma(n+r-1)}\right)^2\overline{z}^{2}\nonumber\\
&&\hspace{-18mm}\times\frac{{_{5}F_{2r+2}\left(\left[2-n,n,n+1,n+r-1,n+r\right],\left[n,n,
..., n+r-1,n+r-1\right],-|z|^{2}\right)}}{_{1}F_{2r-2}\left(\left[-n\right],\left[n,n, ..., n+r-2,n+r-2\right],-|z|^2\right)},\nonumber\\
&&\hspace{-20mm}\left\langle{{K}_{+}}{K}_{-}^{\lambda}\right\rangle=\left(\frac{\Gamma(n+1)}
{\Gamma(n+r-1)}\right)^2|z|^2\nonumber\\
&&\hspace{-18mm}\times\frac{{_{2}F_{2r-1}\left(\left[1-n,1-n\right],\left[-n,n+1,n+1,
...,n+r-1,n+r-1\right],-|z|^{2}\right)}}{_{1}F_{2r-2}\left(\left[-n\right],\left[n,n, ..., n+r-2,n+r-2\right],-|z|^2\right)},\nonumber\\
&&\hspace{-20mm}\left\langle{K}_{3}\right\rangle=-\left(\frac{n}{2}\right)\frac{{_{2}F_{2r-1}\left(\left[-n,1-\frac{n}{2}\right],\left[-\frac{n}{2},n,n,
...,n+r-2,n+r-2\right],-|z|^{2}\right)}}{_{1}F_{2r-2}\left(\left[-n\right],\left[n,n, ..., n+r-2,n+r-2\right],-|z|^2\right)},\nonumber\\
\end{eqnarray}
the squeezing quantities $S^{n}_{1(2)}$ can be easily evaluated for states $\left|z\right\rangle_r$.\\\\
In figure (4), we plot the $S^{n}_{1(2)}$ as a function of $|z|^{2}$
for standard two-mode CSs. As shown in figures 4(a) and (b), the
curves of $S_1$ and $S_2$ corresponding to squeezing in the $X_1$
and $X_2$ operators indicate that for $n=1$, there is a quadrature
squeezing effect in the both $X_1$ and $X_2$ components. From this
figure, squeezing effect occurs in the $X_1 (X_2)$ component for
$\phi=\frac{\pi}{6} (\phi=\frac{\pi}{4})$ in the major range of
$|z|^2$, while for $\phi=0 (\phi=\frac{\pi}{2})$ the squeezing
parameters $S_1 (S_2)$ are negative in all the range of $|z|^2$.
Also, in limit of $|z|^2\rightarrow 1$, maximal squeezing in the
$X_1$ and $X_2$ components occur for $\phi=0$ and
$\phi=\frac{\pi}{2}$, respectively.

 In figure (5), we plot the $S_{1}$ and $S_2$ against $|z|^{2}$ for different
values of $r$(= 2, 3 and 4) as well as $n( =1, 2, 3)$ with respect
to generalized $su(2)$ CSs, $|z\rangle_{r}$. Here, we
choose the phase $\phi=0$. From this figure, the quadrature
squeezing occur only in the $X_1$ component for different values of
$n$. As shown in figure $(a)$, for $n=1$ the squeezing parameter
$S_1$ is negative in all the range of $|z|^2$. This indicate that
for $n=1$ and $\phi=0$, the states $|z\rangle_{r}$ are squeezed
states. Also, we show in Figures (6) the squeezing parameter $S_{1}$
and $S_{2}$ for different values of $\phi$(= 0, $\frac{\pi}{6}$,
$\frac{\pi}{4}$, $\frac{\pi}{3}$, and $\frac{\pi}{2}$) for fixed
$r=3$. This figure show that while quadrature squeezing in the $X_1$
component occur for $\phi=0$ and $\phi=\frac{\pi}{6}$, there is
quadrature squeezing effect in the $X_2$ component for
$\phi=\frac{\pi}{3}$ and
$\phi=\frac{\pi}{2}$.\\

$\diamondsuit${\em{\textbf{Photon statistics}}}\\
Now we are in a position to study the photon statistics of the
constructed generalized $su(2)$ CSs given by Eq. (13). For
this purpose, we calculate the Mandel parameter of bosonic fields in
the first and the second modes \cite{Mandel}
\renewcommand\theequation{\arabic{equation}}
\setcounter{equation}{\value {tempeq}}
\begin{eqnarray}
&&\hspace{-34mm}Q_{i}^{(r)}(|z|^2)= \frac{\langle
{\hat{N_i}}^2\rangle_{r}-{\langle{\hat{N}_i
\rangle}^2_{r}}}{{\langle{\hat{N}_i\rangle}_{r}}}-1,\hspace{1.3cm}
i=(a \,or \,b)
\end{eqnarray}
where, $\hat{N}_i$ is photon number operator in first and second
modes. In fact, a quantum state exhibits super- Poissonian
(photon-bunching), Poissonian and sub-Poissonian
(photon-antibunching) statistics, respectively, if $Q_{i}^{(r)}>0$,
$Q_{i}^{(r)}=0$ and $Q_{i}^{(r)}<0$. Now, evaluating the mean value
of $\hat{N}_i$ and $\hat{N_i}^2$ with respect to the states
$\left|z\right\rangle_r$ yield \setcounter{tempeq}{\value{equation}}
\begin{eqnarray}
&&\hspace{-26mm}{\langle{\hat{N}_b
\rangle}_{r}}=\left(\frac{|z|\sqrt{n}\,\Gamma(n)}{\Gamma(n+r-1)}\right)^{2}\nonumber\\
&&\hspace{-6mm}\times\frac{{_{1}F_{2r-2}\left(\left[1-n\right],\left[n+1,n+1,...,n+r-1,
n+r-1\right],-|z|^{2}\right)}}
{_{1}F_{2r-2}\left(\left[-n\right],\left[n,n,...,n+r-2, n+r-2\right],-|z|^2\right)},\nonumber\\
&&\hspace{-26mm}{\langle{\hat{N_b}^{2}
\rangle}_{r}}=\left(\frac{|z|\sqrt{n}\,\Gamma(n)}{\Gamma(n+r-1)}\right)^{2}\nonumber\\
&&\hspace{-6mm}\times\frac{{_{2}F_{2r-1}\left(\left[2,1-n\right],\left[1,n+1,n+1,...,n+r-1,
n+r-1\right],-|z|^{2}\right)}}
{_{1}F_{2r-2}\left(\left[-n\right],\left[n,n,...,n+r-2,
n+r-2\right],-|z|^2\right)},\nonumber\\
&&\hspace{-26mm}{\langle{\hat{N}_a
\rangle}_{r}}=n-{\langle{\hat{N}_b
\rangle}_{r}},\nonumber\\
&&\hspace{-26mm}{\langle{\hat{N_a}^{2}
\rangle}_{r}}={\langle{\hat{N_b}^{2}\rangle}_{r}}-2n{\langle{\hat{N}_b
\rangle}_{r}}+n^2.
\end{eqnarray}
\begin{figure}
\centering
\includegraphics[width=500 pt]{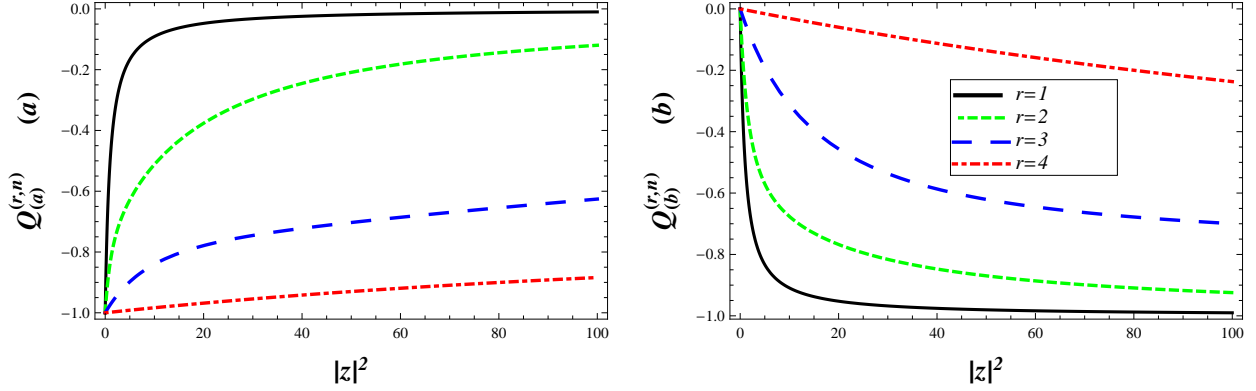}
\caption{Plot of the Mandel parameters (a) $Q_{a}^{(r)}$ and (b)
$Q_{b}^{(r)}$ versus $|z|^2$ for different values of $r$.}
\end{figure}
In Figure 7, the Mandel parameters of two modes, $Q_{a}^{(r)}$ and
$Q_{b}^{(r)}$, have been plotted in terms of $|z|^2$ for $n=2$ and
different values of $r$(=1, 2, 3 and 4). As is evident, both modes
have sub-Poissonian statistics for any values of generalization
parameter $r$. From figure 7(a), in the limit of $|z|^2\rightarrow
0$, the Mandel parameter of first mode is $Q_{a}^{(r)}=-1$, which
indicate that first mode is maximally sub-Poissonian. This situation
is changed for second mode. As shown in figure 7(b), the second mode
is maximally sub-Poissonian for $|z|^2\rightarrow \infty$. Also, by
increasing the generalization parameter $r$, photon statistics of
the first (second) mode of the generalized CSs tends to
sub-Poissonian (Poissonian) more rapidly. In other words,
nonclassical properties of the first (second) mode of the
generalized CSs increases (decreases) by increasing (decreasing) of
generalization parameter $r$.
\begin{figure}
\centering
\includegraphics[width=500 pt]{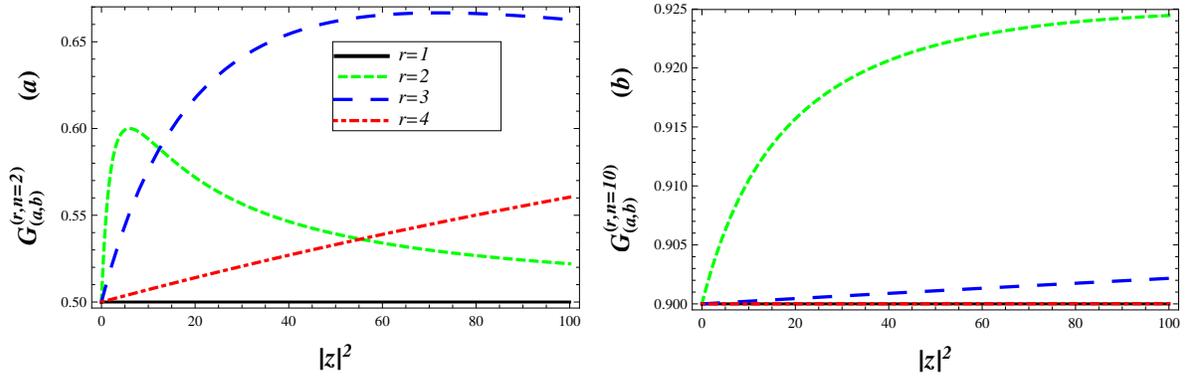}
\caption{Plot of Cross-correlation function $G_{r}^{(2)}$ versus
$|z|^2$ for different values of $r$, with (a) n=2 and (b) n=10.}
\end{figure}
 Now we turn our attention to the anticorrelation properties of the
fields under consideration. Anticorrelations are described by the
normalized cross-correlation functions
$G_{r}^{(2)}$:\begin{eqnarray}
&&\hspace{-14mm}G^{(2)}_{r}(|z|^2)=\frac{\langle{\hat{N_b}\hat{N_a}\rangle}_{r}}{\langle{\hat{N_b}\rangle}_{r}\langle{\hat{N_a}\rangle}_{r}},
\end{eqnarray}
which by using Eq. (34) and the following equation
\begin{eqnarray}
&&\hspace{-14mm}\langle{\hat{N_b}\hat{N_a}\rangle}_{r}=n\langle{\hat{N_b}\rangle}_{r}-\langle{\hat{N_b}^2\rangle}_{r},
\end{eqnarray}
we get
\begin{eqnarray}
&&\hspace{-14mm}G^{(2)}_{r}(|z|^2)=\frac{n\langle{\hat{N_b}\rangle}_{r}-\langle{\hat{N_b}^{2}
\rangle}_{r}}{\langle{\hat{N_b}\rangle}_{r}(n-\langle{\hat{N_b}
\rangle}_{r})}.
\end{eqnarray}
Figure 6 displays the cross-correlation function for the
$|z\rangle_r$, with respect to $|z|^2$ for $n=2, 10$ and $r=(1,2,3$
and $4)$. From figure 8(a), for any value of $r$, cross-correlation
function is less than one. This indicates that the two bosonic modes
which are discussed are anti-correlated. Physically, this means that
there is no tendency for bosons in the different modes to be created
or annihilated simultaneously. Also, as shown in figure 8(b), for
large values of $n$, $G^{(2)}_{r}(|z|^2)$ is equal to one, which
indicates that the bosons in the different modes are created or
annihilated independently.

\section{Conclusions}
Based on a new approach, broad range of generalized
generalized $su(2)$ CSs for problem of a charged particle
moving on an infinite flat band in the presence of a constant
magnetic field are constructed. We have shown that these states are
nonlinear truncated CSs with a spacial nonlinearity
function. Then, the nonclassical properties such as $su(2)$
squeezing, quadrature squeezing and sub-Poissonian statistics for
the introduced states in addition to their anti-correlation have
been reviewed, in detail. It has been shown that, they exhibit
$su(2)$ squeezing in both components $X_1$ and $X_2$. Also, from
Figure 5 the quadrature squeezing only in the angular momentum
component $(p)$ for $(M = \omega =\hbar = 1)$. This situation can be
changed by increasing of external magnetic field $B_{ext}$. It is
easy to show that the squeezing in $p$ component can be transformed
into $x$ component by increasing $B_{ext}$ and vice versa. In fact,
the squeezing effect in the $x$ and $p$ component is flexible by
variation of magnetic field. Our numerical results have explicitly
shown that both modes have sub-Poissonian statistics for any values
of generalization parameter $r$. Furthermore, increasing of $r$
leads to the enhancement of nonclassical properties of the first
mode and causes to diminish nonclassical properties of the second
mode. Generally, the approach presented here can be used to
construct new type generalized $su(2)$ CSs for exactly
solvable models, such as two dimensional harmonic oscillator.


\begin{thebibliography}{99}
\bibitem{Schr¨odinger} E. Schr\"{o}dinger, Ann. Phys. Lpz, 79 (1926) 361.

\bibitem{Glauber1} R. J. Glauber, Phys. Rev, 130 (1963) 2529.

\bibitem{Glauber2} R. J. Glauber, Phys. Rev, 131 (1963) 2766.

\bibitem{Klauder1}J. R. Klauder, J. Math. Phys, 4 (1963) 1055.

\bibitem{Klauder2} J. R. Klauder, J. Math. Phys, 4 (1963) 1058.

\bibitem{Sudarshan} E. C. G. Sudarshan, Phys. Rev. Lett, 10 (1963) 277.

\bibitem{Barut} A. O. Barut and L. Girardello, Commun. Math. Phys, 21
(1971) 41.

\bibitem{Perelomov} A. M. Perelomov, Commun. Math. Phys, 26 (1972) 222.

\bibitem{Gazeau1} J. P. Gazeau and J. R. Klauder, J. Phys. A: Math.
Gen, 32 (1999) 123.

\bibitem{Antoine} J. P. Antoine, J. P. Gazeau, P. Monceau, J. R. Klauder and K. A. Penson, J. Math.
Phys, 2 (2001) 2349.

\bibitem{Fukui} T. Fukui and N. Aizawa, Phys. Lett. A, 180 (1993) 308.

\bibitem{Chenaghlou} A. Chenaghlou and H. Fakhri, Mod. Phys. Lett.
A, 17 (2002) 1701.

\bibitem{Stoler} D. Stoler, Phys. Rev. D, 1 (1970) 3217.

\bibitem{Kimble} H. J. Kimble, M. Dagenais and L. Mandel, Phys. Rev.
Lett, 39 (1977) 691.

\bibitem{Short} R. Short and L. Mandel, Phys. Rev. Lett, 51 (1983) 384.

\bibitem{Teich} M. C. Teich and B. E. A. Saleh, J. Opt. Soc. Am. B, 2
(1985) 275.

\bibitem{Slusher} R. E. Slusher, L. W. Hollberg, B. Yurke, J. C. Mertz and J. F. Valley, Phys. Rev.
Lett, 55 (1985) 2409.

\bibitem{Wu} L-A. Wu , H. J. Kimble, J. L. Hall and H. Wu, Phys. Rev.
Lett, 57 (1986) 2520.

\bibitem{Biedenharn} L. C. Biedenharn, J. Phys. A: Math. Gen. 22
(1989) L873.

\bibitem{Yurke} B. Yurke and D. Stoler, Phys. Rev. Lett, 57 (1986) 13.

\bibitem{Noel} M. W. Noel and C. R. Stroud, Jr. in Coherence and Quantum
Optics VII Plenum, New York, 563 (1996) 564.

\bibitem{Matos1} R. L. de Matos Filho and W. Vogel, Phys. Rev. Lett,
76 (1996) 608.

\bibitem{Monroe} C. Monroe, D. M. Meekhof, B. E. King and D. J. Wineland,
Science, 272 (1996) 1131.

\bibitem{Solomon} A. I. Solomon, Phys. Lett. A 196 (1994) 29; J. Katriel and A. I. Solomon, Phys. Rev. A 49 (1994)
5149; P. Shanta, S. Chaturvdi, V. Srinivasan and R. Jagannathan, J.
Phys. A: Math. Gen. 27 (1994) 6433.

\bibitem{Manko} V. I. Manko, G. Marmo, E. C. G. Sudarshan, F. Zaccaria, Phys.
Scripta. 55 (1997) 528.

\bibitem{Matos2} R. L. de Matos Filho and W. Vogel, Phys. Rev. A, 54
(1996) 4560.

\bibitem{Raffa} F. A. Raffa, M. Rasetti and M. Genovese, Phys. Lett.
A, 376 (2012) 330.

\bibitem{Kis} Z. Kis, W. Vogel and L. Davidovich, Phys. Rev. A, 64
(2001) 033401.

\bibitem{Wang} X. G. Wang, Can. J. Phys, 79(5) (2001) 833.

\bibitem{Tavassoly 2} G. R. Honarasa, M. K. Tavassoly, M. Hatamia, R. Roknizadeh, Physica. A, 390 (2011) 1381; E. Piroozi and M. K.
Tavassoly, J. Phys. A: Math. Theor. 45 (2012) 135301; R. Roknizadeh
and M. K. Tavassoly, J. Phys. A: Math. Gen. 37 (2004) 8111; R.
Roknizadeh and M. K. Tavassoly, J. Math. Phys, 46 (2005) 042110.

\bibitem{Landau} L. Landau, E. Lifchitz, Quantum Mechanics:
Non-Relativistic Theory, Pergamon, New York, 1977.

\bibitem{Laughlin} R. B. Laughlin, Phys. Rev. B, 23 (1981) 5632-5633.

\bibitem{Wybourne} B. G. Wybourne, Rep. Math. Phys, 34 (1994) 9-16.

\bibitem{Chakraborty} T. Chakraborty, P. Pietil\"ainen, The Quantum Hall
Effects - Fractional and Integral, Springer, New York, 1995.

\bibitem{Bagarello1} F. Bagarello, J. Math. Phys, 42 (2001) 5116-5129.

\bibitem{Richter} T. Richter, H. Schulz-Baldes, J. Math. Phys. 42 (2001) 3439-3444.

\bibitem{Bagarello2} F. Bagarello, J. Phys. A: Math. Gen. 36 (2003) 123-138.

\bibitem{Murthy} G. Murthy, R. Shankar, Rev. Mod. Phys, 75 (2003) 1101-1158.

\bibitem{Karabali} D. Karabali, V.P. Nair, J. Phys. A: Math. Gen. 39 (2006) 12735-12763.

\bibitem{Malkin1} I. A. Malkin, V. I. Man'ko, Zh. Eksp. Teor. Fiz, 55
(1968) 1014 [Sov. Phys. JETP 28 (1969) 527].

\bibitem{Feldman} A. Feldman, A. H. Kahn, Phys. Rev. B, 1 (1970)
4584-4589.

\bibitem{Ferreyra} J. M. Ferreyra, C.R. Proetto, J. Phys.: Condens. Matter, 6 (1994) 6623-6636.

\bibitem{Rohringer} N. Rohringer, J. Burgdorfer, N. Macris, J. Phys. A: Math. Gen. 36 (2003) 4173-4190.

\bibitem{Fakhri1} H. Fakhri, J. Phys. A: Math. Gen. 37 (2004) 5203-5210.

\bibitem{Mouayn} Z. Mouayn, Rep. Math. Phys, 55 (2005) 269-276.

\bibitem{Kowalski} K. Kowalski, J. Rembieli\'nski, J. Phys. A: Math. Gen. 38
(2005) 8247-8258.

\bibitem{Yang1} S. J. Yang, Z. Tao, Y. Yu, S. Feng, J. Phys,: Condens. Matter 18 (2006) 11255-11262.

\bibitem{Bracken} P. F. Bracken, Int. J. Theor. Phys,  46 (2007) 119-132.

\bibitem{Yang2} W.-L. Yang, J.-L. Chen, Phys. Rev. A 75 (2007) 024101(4 pages).

\bibitem{Twareque} S. Twareque Ali, F. Bagarello, J. Math. Phys, 49  (2008) 032110(17 pages).

\bibitem{Gazeau2} J. P. Gazeau, M.C. Baldiotti, D.M. Gitman, Phys. Lett. A, 373 (2009)
1916-1920.

\bibitem{Fakhri2} H. Fakhri, B. Mojaveri, M.A. Gomshi Nobary, Rep. Math. Phys, 66 (2010) 299-310.

\bibitem{Mojaveri5} A. Dehghani, H. Fakhri, B. Mojaveri, J. Math. Phys, 53, 123527
(2012).

\bibitem{Mojaveri4} B. Mojaveri, Eur. Phys. J. D, (2013) 67: 105.

\bibitem{Mojaveri7} A. Dehghani and B. Mojaveri, J. Phys. A: Math.
Theor, 46 (2013) 385303.

\bibitem{Dehghani} A. Dehghani and B. Mojaveri, J. Phys. A: Math.
Theor, 45 (2012) 095304.

\bibitem{Mojaveri6} B. Mojaveri and A. Dehghani, Eur. Phys. J. D, (2013) 67: 179.


\bibitem{Buzek3} V. Bu\v{z}ek and T. Quang, J. Opt. Soc. Amer. B, 6 (1989) 2447.


\bibitem{Appl} T. Appl and D. H. Schiller, J. Phys. A: Math. Gen. 37
(2004) 2731.

\bibitem{Gradshteyn} I. S. Gradshteyn and I. M. Ryzhik, Table of Integrals, Series, and Products (San Diego, CA:
Academic) 2000.

\bibitem{Mahdifar1} A. Mahdifar, R. Roknizadeh and M. H. Naderi, J. Phys. A, 39 (2006) 7003.


\bibitem{Mahdifar2} A. Mahdifar, Int. J. Geom. Methods Mod. Phys, 10 (2013) 1350028.

\bibitem{Kuang} L. M. Kuang, F. B. Wang, and Y. G. Zhou, Phys. Lett. A ,183, 1 (1993).

\bibitem{Shchukin} E. V. Shchukin and W. Vogel, Phys. Rev. A, 72 (2005) 043808.

\bibitem{Walls} D. F. Walls, Nature 306 (1983) 141.

\bibitem{WODKIEWICZ} K. Wodkiewwicz and J. H. Eberly J H, J. opt . Soc. Am. B
2 (1985) 458K.

\bibitem{Buzek} V. Bu\v{z}ek, J. Mod. Opt 37 (1990)303.

\bibitem{Mandel} L. Mandel and E. Wolf, Optical Coherence and Quantum Optics (Cambridge University
Press, Cambridge, 1995).

\end{thebibliography}
\end{document}